\documentclass[sigconf]{acmart}

\usepackage{booktabs} 
\usepackage{multirow}

\setcopyright{acmcopyright}

\acmDOI{xx.xxx/xxx_x}

\acmISBN{979-8-4007-0243-3/24/04}

\acmConference[SAC'24]{ACM SAC Conference}{April 8 –April 12, 2024}{Avila, Spain}
\acmYear{2024}
\copyrightyear{2024}

\acmArticle{4}
\acmPrice{15.00}

\begin{document}

\title{Authoring Worked Examples for Java Programming\\ with Human-AI Collaboration}

\titlenote{Produces the permission block, and
  copyright information}
  
\renewcommand{\shorttitle}{SIG Proceedings Paper in LaTeX Format}


\author{Mohammad Hassany}
\email{moh70@pitt.edu}
\orcid{0009-0004-8893-8454}
\affiliation{%
  \institution{University of Pittsburgh}
  \city{Pittsburgh}
  \state{PA}
  \country{USA}
  \postcode{15260}
}

\author{Peter Brusilovsky}
\email{peterb@pitt.edu}
\affiliation{%
  \institution{University of Pittsburgh}
  \city{Pittsburgh}
  \state{PA}
  \country{USA}
  \postcode{15260}
}

\author{Jiaze Ke}
\email{jiazek@andrew.cmu.edu}
\affiliation{%
  \institution{Carnegie Mellon University}
  \city{Pittsburgh}
  \state{PA}
  \country{USA}
  \postcode{15213}
}

\author{Kamil Akhuseyinoglu}
\email{kaa108@pitt.edu}
\affiliation{%
  \institution{University of Pittsburgh}
  \city{Pittsburgh}
  \state{PA}
  \country{USA}
  \postcode{15260}
}

\author{Arun Balajiee Lekshmi Narayanan}
\email{arl122@pitt.edu}
\affiliation{%
  \institution{University of Pittsburgh}
  \city{Pittsburgh}
  \state{PA}
  \country{USA}
  \postcode{15260}
}

\renewcommand{\shortauthors}{}

\begin{abstract}
Worked examples (solutions to typical programming problems presented as a source code in a certain language and are used to explain the topics from a programming class) are among the most popular types of learning content in programming classes. Most approaches and tools for presenting these examples to students are based on line-by-line explanations of the example code. However, instructors rarely have time to provide line-by-line explanations for a large number of examples typically used in a programming class. In this paper, we explore and assess a human-AI collaboration approach to authoring worked examples for Java programming. We introduce an authoring system for creating Java worked examples that generates a starting version of code explanations and presents it to the instructor to edit if necessary. We also present a study that assesses the quality of explanations created with this approach.
\end{abstract}





\maketitle
\section{Introduction}


Program code examples play a crucial role in learning how to program~\citep{linn1992CACM}. Instructors use examples extensively to demonstrate the semantics of the programming language being taught and to highlight the fundamental coding patterns. Programming textbooks also pay a lot of attention to examples, with a considerable textbook space allocated to program examples and associated comments. 

Through this practice, worked code examples emerged as an important type of learning content in programming classes. Following the tradition established by a number of programming textbooks~\citep{deitel1994,kelley1995}, a typical worked example presents a code for solving a specific programming problem and explains the role and function of code lines or code chunks. In textbooks, these explanations are usually presented as comments in the code or as explanations on the margins. While informative, this approach focused on passive learning, which is known for its low efficiency. Recognizing this problem, several research teams developed learning tools that offered more interactive and engaging ways to learn from examples~\citep{brusilovsky2009problem,CODECAST2017,khandwala2018codemotion,park2018elicast,Hosseini2020}. These tools demonstrated their effectiveness in classroom studies, but their practical impact, i.e., broader use by programming instructors was limited due to the \emph{authoring bottleneck}. Although the authors of example-focused learning tools usually provide a good set of worked examples that can be presented through their tools, many instructors prefer to use their own favorite code examples. The instructors are usually happy to broadly share the code of examples they created (usually providing it on the course Web page), but they rarely have time or patience to augment examples with explanations and add their examples to an example-focused interactive system. Indeed, producing a single explained example could take 30 minutes or more, since it requires typing an explanation for each code line~\citep{brusilovsky2009problem,Hosseini2020} or creating a screencast in a specific format~\citep{CODECAST2017,park2018elicast}. 

The authoring bottleneck has been recognized by several research teams, which have offered several ways to address it. Among the approaches explored are learner-sourcing, that is, engaging students in creating and reviewing explanations for instructor-provided code~\citep{hsiao2011role} and automatic extraction of information content from available sources, such as lecture recordings~\citep{khandwala2018codemotion}. In this paper, we present an alternative approach to address the authoring bottleneck based on human-AI collaboration. With this approach, the instructor provides the code of one of their favorite examples along with the statement of the programming problem it is solving. The AI engine based on large language models (LLM) examines the code and generates explanations for each code line. The explanations could be reviewed and edited by the instructor. To support and explore this authoring approach, we created an authoring system, which radically decreases the time to create a new interactive worked example. The examples created by the system could be uploaded to an example-exploration system such as WebEx~\citep{brusilovsky2009problem} or PCEX~\citep{Hosseini2020} or exported in a reusable format. To assess the quality of the resulting examples, we performed a user study in which TAs and students compared code explanations created by experts through a traditional process with examples created by AI to 
contribute to human-AI collaborative process.

The remainder of the paper is structured as following. We start by reviewing related work, introduce the example authoring system that implements the proposed collaborative approach, and explain how specific design decisions were made through several rounds of internal evaluation. Next, we explain the design of our user study and review its results. We conclude with a summary of the work and plans for future research.

\section{Related Work}

\subsection {Worked Examples in Programming}
Code examples are important pedagogical tools for learning programming. Not surprisingly, considerable efforts have been devoted to the development of learning materials and tools to support students in studying code examples.
Hosseini~\citep{Hosseini2020} classified program examples that have been used in teaching and learning to program into two groups, according to their primary instructional goal: \emph{program behavior examples} and \emph{program construction examples}. Program behavior examples are used to demonstrate the semantics (i.e., behavior) of various programming constructs (i.e., what is happening inside a program or an algorithm when it is executed). Program construction examples attempt to communicate important programming patterns and practices by demonstrating the construction of a program that achieves various meaningful purposes.

Program behavior examples have been extensively studied. While textbooks still explain program behavior by using textual comments attached to lines of program code, a more advanced method for this purpose --- \textit{program visualization}, which visually illustrates the runtime behavior of computer programs --- is now considered as state-of-the-art. Over the past three decades, a number of specialized educational tools for observing and exploring program execution in a visual form have been built and assessed \citep{sorva2013review}. 

Computer-based technologies for presenting program construction examples are less explored. For many years, the state-of-the-art approach for presenting worked code examples in online tools was simply code text with comments \citep{linn1992CACM,davidovic2003learning,morrison2016subgoals}. More recently, this approach has been enhanced with multimedia by adding audio narrations to explain the code \citep{ericson2015analysis} or by showing video fragments of code screencasts with the instructor's narration being heard while watching code in slides or an editor window \citep{CODECAST2017,khandwala2018codemotion}. Both ways, however, support \emph{passive} learning, which is the least efficient approach from the prospect of the ICAP framework~\citep{chi2018icap}\footnote{The ICAP framework differentiates four modes of engagement, behaviorially exhibited by learners: \emph{passive, active, constructive and interactive}.}

An attempt to make learning from program construction examples \emph{active} was made in the WebEx system, which allowed students to interactively explore instructor-provided line-by-line comments for program examples via a web-based interface \citep{brusilovsky2009problem}. More recently, several projects~\citep{khandwala2018codemotion,park2018elicast,Hosseini2020} augmented examples with simple problems and other constructive activities to elevate the example study process to the \emph{interactive and constructive} levels of the ICAP framework, known as the most pedagogically efficient. 

\begin{figure}
    \centering
    \includegraphics[width=1\linewidth]{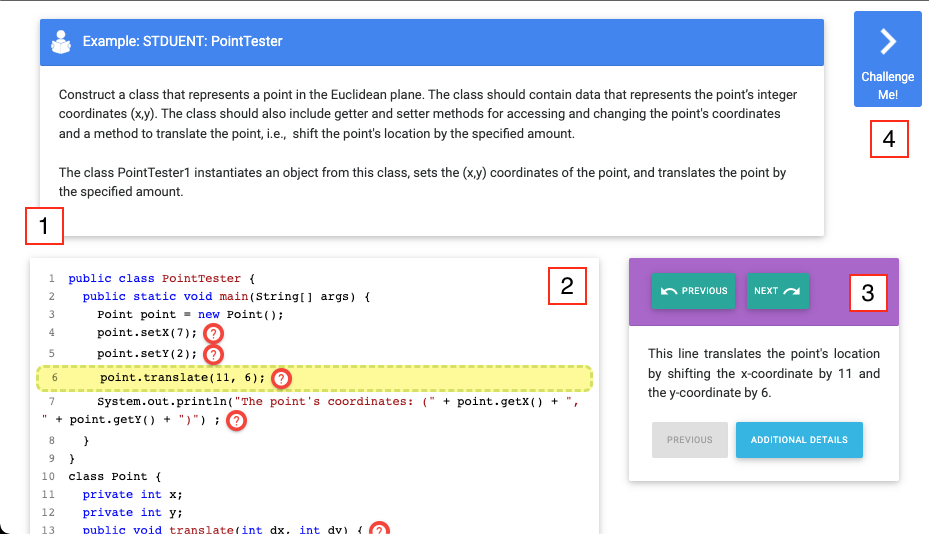}
    \caption{Studying a code example in the PCEX system: 1) title and program description, 2) program source code with lines annotated with explanations, 3) explanations for the highlighted line, 4) link to a ``challenge'' - a small problem related to the example.}
    \label{fig:pcex-example}
\end{figure}

A good example of a modern interactive tool for studying code examples is the PCEX system~\cite{Hosseini2020}. PCEX (Program Construction EXamples) was created in the context of an NSF Infrastructure project (https://cssplice.org) with a focus on broad reuse and has been used by several universities in the US and Europe in the context of Java, Python, and SQL courses.
PCEX interface (Figure~\ref{fig:pcex-example}) provides interactive access to traditionally organized worked examples, i.e., code lines augmented with instructor's explanations. Separating explanations (Figure~\ref{fig:pcex-example}-3) from the code (Figure~\ref{fig:pcex-example}-2), allows students to selectively study explanations for code lines they want. Explanations are provided on several levels of detail, so more details could be requested if the brief explanation is not sufficient (Figure~\ref{fig:pcex-example}-3). 

Since line-by-line multi-level example explanations offered by PCEX is currently the most detailed approach for explaining worked examples, we selected the code example structure implemented by PCEX as the target model for our authoring tool introduced in the next section. The tool produces code augmented with line-by-line explanations on several levels of detail. The resulting example could be directly uploaded to PCEX or exported in a system-independent format to be uploaded to other example exploration systems like WebEx\citep{brusilovsky2009problem}.

\subsection{Use of LLMs for Code Explanations}


Multiple researchers have explored code summarization~\cite{Phillips2022ImprovedEO} and explanations using transformer models~\cite{10113620, peng-etal-2022-rethinking}, abstract syntax trees~\cite{Shi2022CodeDKTAC}, and Tree-LSTM~\cite{tian2023chatgpt}. With the announcement of ChatGPT, several research teams explored the use of LLM for code explanations using GPT 3~\cite{10.1145/3544548.3581388,10.1145/3545945.3569785,leinonen2023comparing}, GPT 3.5~\cite{10.1145/3545945.3569785,li2023explaining,10.1007/978-3-031-36336-8_50}, GPT 4~\cite{li2023explaining}, OpenAI Codex \cite{10.1145/3501385.3543957,tian2023chatgpt,10.1145/3545945.3569785}, and GitHub Copilot \cite{10.1007/978-3-031-36336-8_50}. Table \ref{tab:llm_prior_works} presents a brief summary of the most important prior work.

In the prior work, LLMs were used to generate explanations at different levels of abstraction (line-by-line, step-by-step, and high-level summary). Sarsa et al.~\cite{10.1145/3501385.3543957} observed that ChatGPT can generate better explanations at low-level (lines). Li et al.~\cite{li2023explaining} used the result of specific-to-general generated explanations as one of the inputs to their LLM solver, trying to solve competitive-level programming problems more efficiently. 
A novel research~\cite{10.1145/3544548.3581388} tried to understand how non-expert approach LLMs. They have identified common mistakes and provided advice for tool designers.

Explanations and summaries generated by these LLMs were mostly evaluated by authors~\cite{10.1145/3501385.3543957}, students~\cite{10.1145/3545945.3569785,leinonen2023comparing}, and tool users~\cite{10.1007/978-3-031-36336-8_50}. Sarsa et al.~\cite{10.1145/3501385.3543957} reported a high correct ratio for generated explanations with minor mistakes that can be resolved by the instructor or teaching assistant. Students rated LLM-generated explanations as being useful, easier, and more accurate than learner--sourced explanations~\cite{leinonen2023comparing}. 

Prompt, as an essential part of communication, directly influences the LLM's performance. A verbose prompt will limit the LLM's ability to utilize its knowledge~\cite{tian2023chatgpt}. Iterative prompts are proven to perform well~\cite{10.1145/3544548.3581388}. In terms of code explanation, providing the source code and expected outcome is essential. Adding input/output examples can help generate better explanations. Although LLMs like ChatGPT can understand the natural language very well, researchers suggested writing the prompt as writing a code: following a structure and marking different parts of the prompt~\cite{10.1145/3544548.3581388}. If possible, it is better to control the randomness of LLMs responses (for instance, adjusting the temperature to a lower value, perhaps 0). A temptation to allow non-expert form input prompts will not be any good, as Zamfirescu-Pereira and colleagues~\cite{10.1145/3544548.3581388} observed non-experts have misconceptions about LLMs and will struggle to come up with a well-formed prompt. Researchers believe that LLMs can be beneficial in environments where humans and AI can work together, where the human can perform the expert evaluation and tune the responses generated by the AI while the AI performs the time-consuming manual tasks \citep{white2023prompt}.

\begin{table*}
\centering
\begin{tabular}{|c|p{5cm}|p{1.5cm}|p{4cm}|p{4cm}|}
\hline
Source & Goal & LLM(s) & Type of Explanations  & Evaluation \\ \hline
\cite{10.1007/978-3-031-36336-8_50} & Provide explanations for a code fragment selected in the IDE& GPT 3.5 & Explain the selected code& Interview with students, teachers, and bootcamp tutors\\
\cite{leinonen2023comparing} & Scaffold student's ability to understand and explain code & GPT 3  & Explain the intended purpose of a function & Compare ChatGPT explanations with student/peer explanations\\
\cite{li2023explaining} & Given the problem description and expert solution, ChatGPT is prompted to generate explanations& GPT 3.5 vs GPT 4 & Program summary, used algorithm, step-by-step solution description, time complexity, etc & Generated explanations were evaluated by the human programming expert who authored the ``oracle'' solution \\
\cite{10.1145/3545945.3569785} &Generate specific explanation, summary, and concepts for a given code snippet& GPT 3 and Codex & line-by-line explanations, list of important concepts, high-level summary of the code & Students' ratings of explanations, and their utility time/count \\
\cite{10.1145/3501385.3543957} & Help introductory programming course teachers by creating programming exercises + test cases, and code explanations& Codex & step-by-step explanation, problem-statement-like description, high-level description & Internal evaluation, measuring the percentage of code being explained\\
\hline
\end{tabular}
\caption{Prior works in using LLMs (ChatGPT/Codex) to generate code explanations.}
\label{tab:llm_prior_works}
\end{table*}



\section{Human-AI Collaborative Authoring}

\begin{figure}[h]
    \centering
    \includegraphics[width=1\linewidth]{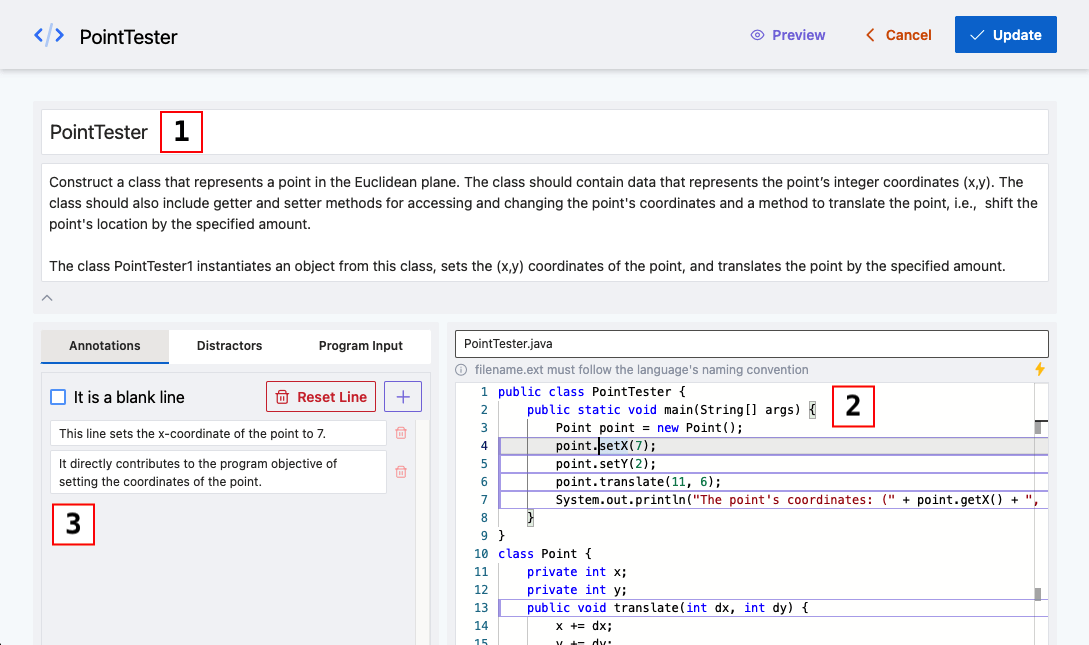}
    \caption{WEAT Authoring, 1) program title and description, 2) program source code (lines with explanations are marked with purple border), 3) explanations for the selected line (line 7 in the screenshot).}
    \label{fig:pcex-authoring}
\end{figure}

Creating a new worked example for an interactive example-focused system such as PCEX is a time-consuming task even in systems that provide some authoring support. Practical instructors who need to create code examples rarely have this time, which results in the authoring bottleneck mentioned above. Our Worked Example Authoring Tool (WEAT) attempts to reduce this bottleneck by engaging ChatGPT in the human-AI collaborative authoring process. In this collaboration, the main task of a human author is to provide the code of the example and the statement of the problem that the code solves. The main task of ChatGPT is to generate the bulk of code line explanations on several levels of detail. As an option, a human author could edit and refine the text produced by ChatGPT to adapt it to the class goals and target students. As in any productive collaboration, each side does what it is best suited to do, leaving the challenging work to the partner. In the main part of the WEAT interface the problem  (Figure \ref{fig:pcex-authoring}-1) and the code (Figure \ref{fig:pcex-authoring}-2) have to be provided by the instructor, while the explanations (Figure \ref{fig:pcex-authoring}-3) are generated by ChatGPT. All generated explanations could be edited by the instructor, who could also turn a regular example into a challenge by marking some lines as blank (within a challenge, these lines can be replaced with distractors, one-line code snippets, by the student). Note that the WEAT interface allows the author to completely replace generated explanations or even create the whole example from scratch, without the help of AI, however, we do not expect that this option will be used frequently.


\begin{figure}
    \centering
    \includegraphics[width=1\linewidth]{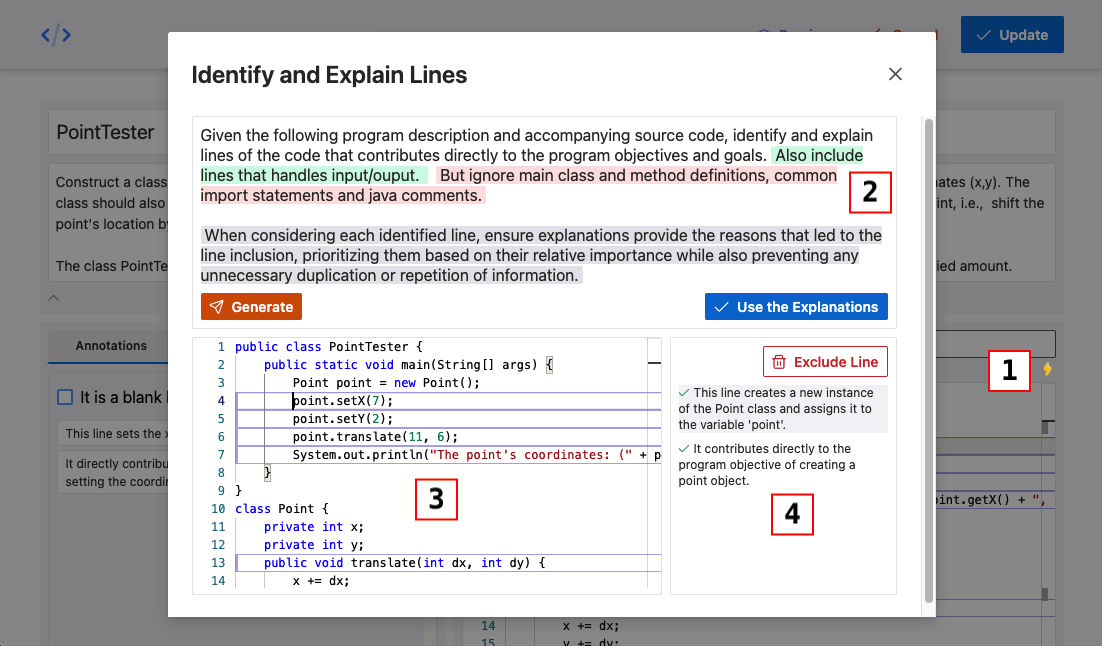}
    \caption{Human-AI Collaborative Worked Example Authoring, 1) open dialog button (small-yellow-bolt icon), 2) default prompt (author can tune the prompt - optional), 3) program source preview (lines with explanations are marked with a purple border), 4) generated explanations for the selected line.}
    \label{fig:human-ai_pcex-authoring}
\end{figure}

To generate ChatGPT explanations for the provided example code and problem description, the author has to click the small-yellow-bolt icon to open the ChatGPT dialog (Figure \ref{fig:human-ai_pcex-authoring}). In this dialog, the author should click "Generate" to generate the explanations, and then click "Use the Explanations" to add them to the example. The WEAT provides the human author several opportunities to control the outcome of the explanation generation process: 1) the author can tune the prompt to their needs, 2) the author can decide whether to include or exclude a generated explanation and 3) the author can edit or remove the explanation after it is edited.

\section{Prompt Tuning through Internal Evaluation}

Following the majority recent work on generating code explanations, we choose ChatGPT as the target LLM to generate code explanations. ChatGPT provides an easy-to-use API and an affordable pricing model. Adding ChatGPT to an application is not a straightforward process and requires careful planning. The key part of this process is crafting a prompt, which requires multiple trials. Following the suggestions in the previous work~\cite{Zhou2022LeasttoMostPE,10.1007/978-3-031-36336-8_50}, the authors used an internal evaluation process to engineer a prompt that produces high-quality explanations. 

To shorten the prompt design process, we adopted several design decisions that were shown to be effective in previous work: assigning a role to ChatGPT \cite{white2023prompt}, avoiding verbosity \cite{tian2023chatgpt}, repetition \cite{10.1145/3544548.3581388}, prompt that looks like code \cite{10.1145/3544548.3581388}, and defining the expected output format \cite{10.1145/3544548.3581388}. However, a few design decisions not evaluated previously were not evident, so we had to use an internal evaluation process to select the best-performing option. The questions answered through the evaluation included the following: 1) Does the presence of a program description in the prompt result in better explanations? 2) Does iterative prompting perform better than a single prompt, and if so, how many iterations are sufficient to have a good explanation? 3) Does adding line inclusion/exclusion criteria in the prompt help ChatGPT to select or ignore lines in generating an explanation? To answer these questions, we formally compared ChatGPT-generated explanations through an independent rating performed by three authors of the paper.

\begin{figure}
    \centering
    \includegraphics[width=0.95\linewidth]{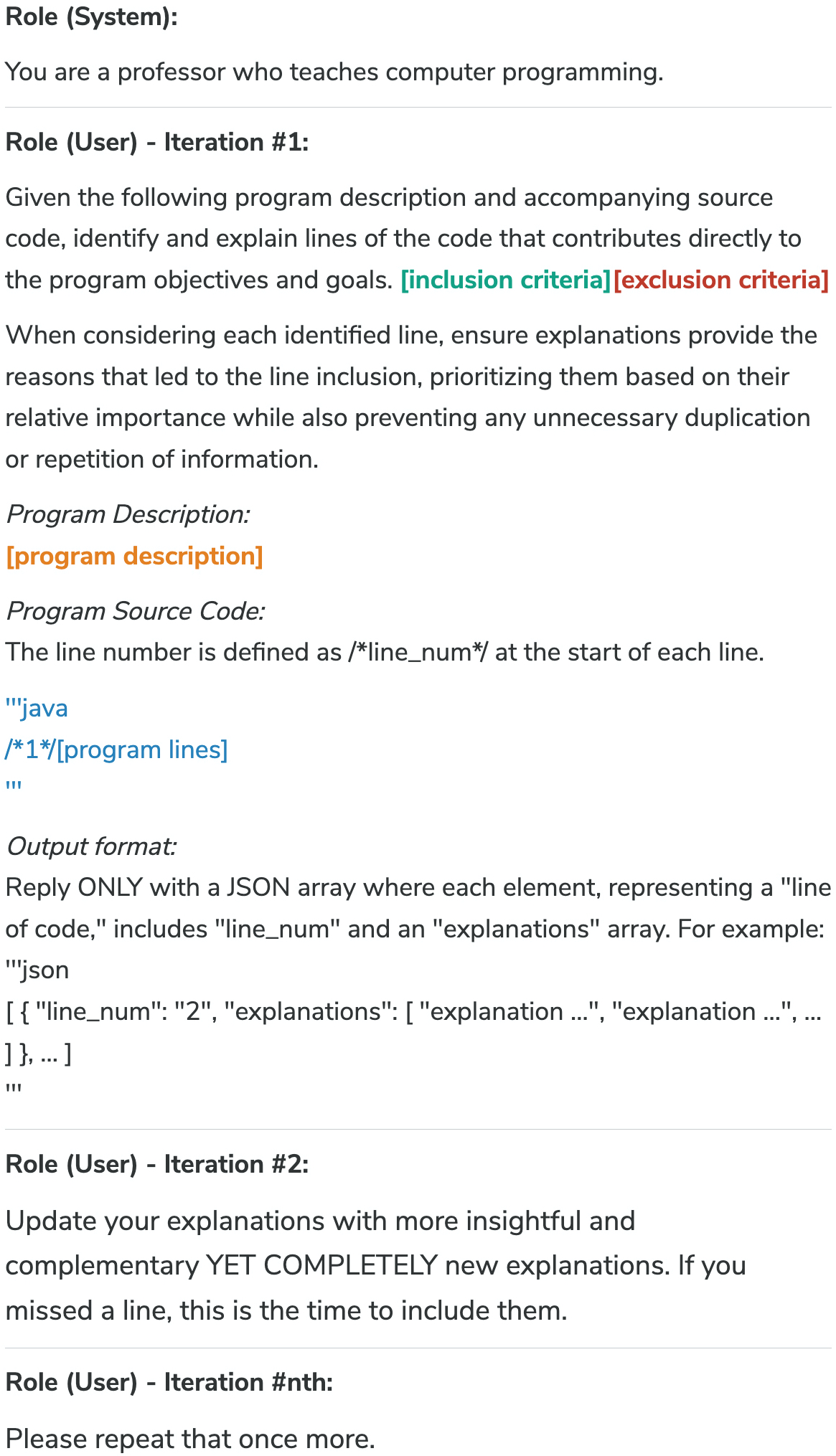}
    \caption{ChatGPT Prompt Template. ChatGPT (is given the "professor" role) is prompted iteratively.}
    \label{fig:prompt-template}
\end{figure}

Since we started from previously explored prompting techniques, the first version of our prompt was reasonably close to our final prompt. At the first stage of the process we made a few small corrections of the prompt based on observations. First, we observed that ChatGPT cannot associate the line number with the line correctly. To address this issue, we marked each line with its line number. We also observed that sometimes with iterative prompting ChatGPT generates duplicate explanations. Hence in our iterative prompts, we asked ChatGPT to generate explanations that are new. Figure~\ref{fig:prompt-template} shows the final version of the prompt that we used with \emph{ChatGPT gpt-3.5-turbo/16k} model (\emph{temperature=0}) through OpenAI API for our internal and external evaluations. 

\emph{Selecting Examples for Evaluation}: We randomly selected eight Java examples with different difficulty levels (string operation, array, loop, and object-oriented programming) from the PCEX repository for the study. Selected examples include:
\begin{itemize}
    \item \emph{Initials}: Extracting initials from full name. 
    \item \emph{JAdjacentDuplicates}: Checks whether a sequence of numbers contains adjacent duplicates.
    \item \emph{JArrayIncrementElements}: Increments all elements of the array by 1.
    \item \emph{JArrayMax}: Finds the maximum value in an array.
    \item \emph{JPrintDigitsReverse}: Prints the digits of an integer from right to left.
    \item \emph{JSearchArrayValues}: Search for values from one array in another.
    \item \emph{JSmallestDivisor}: Smallest divisor of a positive number.
    \item \emph{PointTester}: Translate 2-dimensional coordinates.
\end{itemize}



\begin{table}[h!]
\begin{tabular}{@{}lccc@{}}
\toprule
\textbf{} & \multicolumn{2}{c}{with desc} & \multicolumn{1}{c}{without desc} \\ \midrule
\textbf{Examples} & \(R_2\) & \(R_3\) & \(R_2\) \\ \hline
Initials & 88.8\% & 96.0\% & 90.7\% \\
JAdjacentDuplicates & 93.6\% & 99.0\% & 86.7\% \\
JArrayIncrementElements & 40.0\% & --  & 93.3\% \\
JArrayMax & 85.7\% & -- & 83.8\% \\
JPrintDigitsReverse & 57.1\% & -- & -- \\
JSearchArrayValues & 90.0\% & 71.4\% & 93.0\% \\
JSmallestDivisor & 46.3\% & -- & 86.3\% \\
PointTester & 91.6\% & -- & 73.3\% \\
\bottomrule
\end{tabular}
\caption{Cosine similarity between rounds of explanations (\(R_{n=2} = cosine\_sim(R_n, R_{n-1})\)) with and without including program description.}
\label{tab:cosine-similarity}
\end{table}

\emph{Including/Excluding Program Description}: We hypothesized that adding a program description for the prompt adds information for ChatGPT to produce better explanations, but we were also concerned that it could confuse ChatGPT. To compare the quality of the explanation with and without description, the evaluators checked the explanations for the following: 1) correctness, 2) relevance to the given program description (when present), 3) presence of new information in the $2^{nd}$ round compared to the \(1^{st}\) round, 4) presence of hallucinations when the program description is not present, 5)  whether the $2^{nd}$ round with program description in the prompt had more information than without description. Both Correctness and Relevance were binary ratings. For example,  given an explanation \emph{This line initializes a variable 'fullName' and assigns it the value 'John Smith'.
The 'fullName' variable stores the full name of the person whose initials are to be printed.} for the line of code \emph{String fullName = "John Smith";} was rated as ``correct'' and ``relevant'' by one rater.

\begin{figure}
    \centering
    \includegraphics[width=1\linewidth]{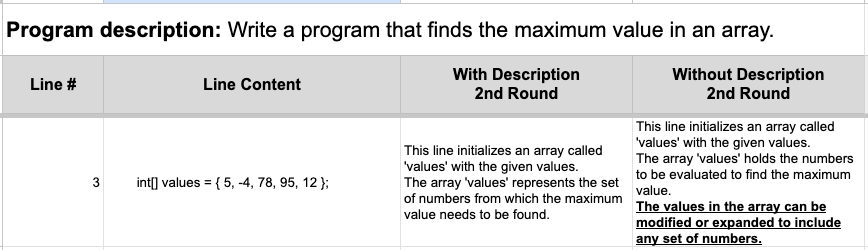}
      \caption{Example of line explanations which the $2^{nd}$ round of explanation when the program description is not present had additional information compared to when present.}
    \label{fig:2nd-round_withoutdesc_gt_withdesc}
\end{figure}

We, as internal evaluators, rated higher correctness in explanations for both rounds 1 and 2 of generation by ChatGPT (\(R_1=99.23\%, R_2=98.77\%\)) as summarized in the Table~\ref{tab:internal-evals_whenpresent}. As an interesting example,  when observing the ratings that we used to compare the amount of new information generated in round 2 from round 1,  we observed that more information is generated without program description (\(48.24\%\)) than with description ((\(35.80\%\)) as summarized in Table~\ref{tab:internal-evals_whennotpresent}. The generated explanations when the program description is not present had additional information compared to when it is present as shown in Figure \ref{fig:2nd-round_withoutdesc_gt_withdesc}. This validates prior work that comprehensive prompts limit LLM's ability to utilize their knowledge \cite{tian2023chatgpt}. Additionally, when the program description is present, the authors selected the $2^{nd}$ round of explanations for the external evaluation because students relate better with the program and it is also rated higher for correctness. In the conditions that we did not include the program description in prompts, we were interested to know to what extent ChatGPT may hallucinate. We observed hallucinations \(2.94\%\ on\ average \) when considering prompts without program description, which could be attributed to greater information generation in round 2. Given this tendency to hallucinate when generating explanations with prompts that exclude problem descriptions, we considered using the explanations that are generated with prompts that include program descriptions.


\begin{table}[]
\begin{tabular}{@{}cccccc@{}}

\toprule
\multicolumn{1}{l}{} & \multicolumn{2}{c}{Round 1} & \multicolumn{3}{c}{Round 2} \\ \midrule
\multicolumn{1}{l}{} & *C & **R & *C & **R & ***A \\
Min & 98.46\% & 44.62\% & 97.53\% & 37.04\% & 32.10\% \\
Max & 100.00\% & 70.77\% & 100.00\% & 68.75\% & 39.51\% \\
Average* & 99.23\% & 55.38\% & 98.77\% & 46.91\% & 35.80\% \\ \bottomrule
\end{tabular}
\caption{Internal evaluators rating, when program description is present in the prompt, *Correctness, **Relevance to program description, ***Additional information compared to 1st round.}
\label{tab:internal-evals_whenpresent}
\end{table}

\begin{table}[]
\begin{tabular}{@{}crccccc@{}}
\toprule
\multicolumn{1}{l}{} & \multicolumn{2}{c}{Round 1} & \multicolumn{3}{c}{Round 2} \\ \midrule
\multicolumn{1}{l}{} & *C & **H & *C & **H & ***A \\
Min & 93.98\% & 0.00\% & 92.94\% & 0.00\% & 41.18\% \\
Max & 100.00\% & 4.82\% & 100.00\% & 5.88\% & 55.29\% \\
Average* & 96.99\% & 2.41\% & 96.47\% & 2.94\% & 48.24\% \\\bottomrule
\end{tabular}

\caption{Internal evaluators rating, when program description is not present in the prompt, *Correctness, **Explanation contains hallucinations, ***Additional information compared to 1st round.}
\label{tab:internal-evals_whennotpresent}
\end{table}

\emph{Assessing Multi-Round Prompting}: In this step, we assessed whether iterative prompting ChatGPT to provide an explanation (see Figure~\ref{fig:prompt-template}) results in additional explanations. When the program description was present in the prompt, only 3 out of 9 examples had additional explanations compared to none when not included (Table \ref{tab:cosine-similarity}). Explanations generated in the \(3^{rd}\) round were either minor wording changes (high cosine similarity) or included explanations for unnecessary lines (closing bracket for main method and class). Qualitatively assessing explanations generated in the \(2^{nd}\) round, they included additional explanations or improved wordings. The number of additional explanations or improvements was not consistent among the examples in the \(2^{nd}\) round, but on average, in \(35.80\%\) of lines (Figure \ref{tab:internal-evals_whenpresent}) when the program description is present in the prompt, and in \(48.24\%\) of lines (Figure \ref{tab:internal-evals_whennotpresent}) when not, additional information was reported by the evaluators. Based on these findings, we decided to adopt a two-round prompting option for WEAT and used this option in the external evaluation process. We summarize our results in Table~\ref{tab:cosine-similarity}. 

\emph{Assessing Inclusion/Exclusion Criteria}: A program description can provide a rich context for identifying and explaining lines of code. However, ChatGPT may sometimes include an unnecessary line or exclude a necessary one from the explanation. Initially, we assumed that directly adding inclusion/exclusion criteria in the ChatGPT prompt can address this issue. However, evaluating this option internally, the authors observed that it resulted in less than \(1\%\) new lines inclusion and around \(4-6\%\) of lines exclusion. When these criteria are present in the prompt, ChatGPT ends up having unnecessary rounds of explanations. Sometimes ChatGPT falls into a loop where it flips wordings between each round. Since the author can review and ignore the explanations for a specific line in the authoring interface, we decided not to use Inclusion/Exclusion criteria in the prompt.

\section{Evaluation}

To assess the quality of the explanations generated by ChatGPT using the best-performing prompt with options tuned through the internal evaluation, we performed a user study.
In this study, we compared the explanations generated by ChatGPT with explanations created by experts for the same PCEX examples. 
Unlike some earlier studies that used beginner students to evaluate ChatGPT explanations, we used more experienced users -- advanced undergraduate and graduate students. The reason for this difference is that in our authoring system, the direct users of the ChatGPT explanation are not \emph{consumers} of explanations, but prospective \emph{authors}. In the implemented human-AI collaborative authoring approach, authors have the option to edit the generated explanation. Thus, it is up to the prospective authors to decide how good the explanations are since their perception of quality impacts the amount of their work: poor explanations will require a lot of editing, while good explanations could be accepted as-is or with minimal changes.

To support these evaluation needs, we recruited 15 evaluators, of which 6 were graduate students doing research on computing education and 9 were undergraduate students who just completed an advanced Java programming class. Graduate students selected for the study usually serve as assistants or instructors in programming classes where supplementary content development is their major responsibility. For brevity, we refer to them as \emph{authors} in our analysis. Advanced undergraduate students are frequently involved in learning content production through ``learnersourcing''~\citep{hsiao2011role,williams2016axis}. To distinguish them from the true authors, we refer to them as \emph{students}. 
Participants had to provide their responses through an evaluation form. The evaluation was estimated to take one hour to complete. Participants received an Amazon gift card of US \$20 as compensation. 

The evaluation form included 8 examples introduced above. 
For each example, the form included a program description and the example code. For each line of each code example, it listed an explanation generated for this line by ChatGPT and by an expert. 
The participants had to rate both explanations for a given line of code and compare them. The order of ChatGPT and expert explanations for a given line of code was randomized, and the evaluators did not know which explanation was generated by ChatGPT or the expert. 
``Expert'' explanations were extracted from real worked examples in PCEX system~\cite{10.1145/3279720.3279726}. These explanations were authored by instructors and TAs and polished through several years of classroom use.

To evaluate the explanations, the participants had to rate to what extent each explanation is \emph{complete} and which is \emph{better}. We defined a \emph{better} explanation as ``providing more information, going deeper, better connecting to programming concepts''. However, we did not provide the definition to a \emph{complete} explanation. 

More specifically, participants had to rate the two explanations with the following metrics:
\begin{enumerate}
    \item \emph{Explanation 1 is sufficiently complete}: Not complete (0), Complete (1), Very complete (2)
    \item \emph{Explanation 2 is sufficiently complete}: Not complete (0), Complete (1), Very complete (2)
    \item \emph{Which explanation is better?} Both are the same (0), Explanation 1 is better (1), Explanation 2 is better (2) 
\end{enumerate}



From the collected responses, we excluded lines that had either ChatGPT or expert explanations but not both. In these cases, the evaluators generally rated the explanations as better without comparison with a missing counterpart explanation. Altogether, there were 18 lines that were explained by ChatGPT but not by the expert, and 5 lines explained by the expert but not ChatGPT. Looking closer at these lines, we observed that 4 out of 5 missing lines by ChatGPT were in the \emph{PointTester} example, which included class definition, object instantiation, and instance variable definition. We are not aware of the reason why the expert didn't explain these lines, but we assume these lines are either mentioned in explanations generated for other lines or they don't provide important information toward understanding the program. Although the program description had related wordings, there were missed by ChatGPT: ``\emph{Construct a class} that represents... The class should contain \emph{data that represents the point’s integer coordinates(x, y)}. ... The class \emph{PointTester instantiates an object from this class}, sets the (x, y) coordinates of the ...''. Conversely, in 14 out of 18 of these lines, ChatGPT unnecessarily explained class, main method definition, and closing brackets (class, method, loop, and condition). The other 4 lines were informative and useful. This can support the importance of having inclusion criteria in the prompt. 

For the remaining 45 lines of code, we observed from the evaluators' ratings for the question ``Explanation 1 is sufficiently complete?'' or ``Explanation 2 is sufficiently complete?'' that ChatGPT explanations were rated as \(0.59\%\) (not complete), \(21.04\%\) (complete) and \(78.37\%\) (very complete) compared to Expert explanations as \(6.96\%\) (not complete), \(56.44\%\) (complete), and \(36.59\%\) (very complete). In response to the question ``Which explanation is better?'', evaluators selected ChatGPT as the better explanation in \(53.93\%\) of lines, compared to experts (\(20.59\%\)); and in the rest of the lines (\(25.48\%\)) both were rated as the same. Our calculations of the inter-rater reliability for the ratings of the question ``Which explanation is better?'' using Fleiss-Kappa gave us \(0.182,\ p < 0.01\) score of agreement. This can be interpreted as "slight agreement" based on the 2-raters/2-categories table. Given that Fleiss-Kappa is a chance-corrected coefficient, it can be interpreted as a better agreement due to the high number of subjects (45 lines of code by 15 evaluators) ~\cite{10.1093/ptj/85.3.257}.

We observe that the students did not rate ChatGPT explanations incomplete at all with their \(13.33\%\) and \(86.67\%\) ratings being that ChatGPT explanations are complete and very complete respectively. Authors also rated ChatGPT explanations as complete (\(32.59\%\)) or very complete (\(65.93\%\)). Hence, a majority of authors and students find ChatGPT explanations complete, as shown in Table~\ref{tab:chatgpt-expert_completeness}. In terms of comparing the explanations for which is better, \(51.11\%\) and \(58.15\%\) of students and authors, respectively, find that the explanations of ChatGPT are better for the given lines of code. On average, the authors rated the ChatGPT explanations more complete than students and students preferred ChatGPT explanations more than the authors, as summarized in Table~\ref{tab:avgstdev}. A direct comparisons of two options, based on the question "which explanation is better (ChatGPT vs Expert)?", is presented in Table~\ref{tab:chatgpt-expert_whichisbetter}.  Given that the assessment was performed using blind rating, this is an encouraging result for the use of generative AI for authoring tools. 


\begin{table}[t]
\begin{tabular}{@{}lccc@{}}
\toprule
& Not complete=0 & Complete=1 & Very complete=2 \\ \midrule
ChatGPT &  &  &  \\
\ \ \ Students & 0.00\% & 13.33\% & 86.67\% \\
\ \ \ Authors & 1.48\% & 32.59\% & 65.93\% \\ 
\ \ \ Overall & 0.59\% & 21.04\% & 78.37\% \\
\hline
Expert &  &  &  \\
\ \ \ Students & 2.22\% & 55.56\% & 42.22\% \\
\ \ \ Authors & 14.07\% & 57.78\% & 28.15\% \\ 
\ \ \ Overall & 6.96\% & 56.44\% & 36.59\% \\
\bottomrule
\end{tabular}
\caption{Percentage of Ratings for different items on the scale for ``Explanation 1 / 2 is sufficiently complete?''}
\label{tab:chatgpt-expert_completeness}
\end{table}



\begin{table}[t]
\begin{tabular}{@{}lccc@{}}
\toprule
\multicolumn{1}{l}{} & \multicolumn{3}{c}{Explanation} \\
\cmidrule{2-4}
Rating & Students & Authors & Overall \\
\hline
Both are the same = 0 & 32.84\% & 14.44\% & 25.48\% \\
Expert is better = 1 & 16.05\% & 27.41\% & 20.59\% \\
ChatGPT is better = 2 & 51.11\% & 58.15\% & 53.93\% \\
\bottomrule
\end{tabular}
\caption{Percentage of Ratings for the different items on the scale for ``Which explanation is better?''}
\label{tab:chatgpt-expert_whichisbetter}
\end{table}

\begin{table}[]
\begin{tabular}{@{}cccc@{}}
\toprule
\multicolumn{1}{l}{} & All & Students & Authors \\ \midrule
ChatGPT* & 1.867 (0.133) & 1.644 (0.258) & 1.778 (0.163) \\
Expert* & 1.400 (0.388) & 1.141 (0.465) & 1.296 (0.408) \\
Which is better? & 1.183 (0.510) & 1.437 (0.373) & 1.284 (0.427) \\ \bottomrule
\end{tabular}
\caption{Average (Stdev) Ratings - *Completeness}
\label{tab:avgstdev}
\end{table}


\section{Conclusion}

In this paper, we introduce a worked example authoring tool that utilizes ChatGPT for the automatic generation of line-by-line code explanations. The tool is designed to allow human and AI to collaborate in the process of authoring such examples. To the best of our knowledge, this is the first attempt to produce worked example through human-AI collaboration. Our work supports findings by other researchers and provides empirical evidence on the value of using ChatGPT for generating line-by-line code explanations. Through an external evaluation, this work also compared the generated explanations and human expert explanations. 

As the first step towards this important goal, our work has several limitations. First, the scale of our evaluation was relatively small. Since we targeted prospective authors as users in our evaluation process, we were able to recruit only 15 qualified subjects. Furthermore, within the time allocated for the study, the subjects were able to process only eight worked examples. Although we attempted to broadly vary the topics and difficulty of selected examples to achieve sufficient generalizability of results, a larger-scale study with a broader variety of examples might be necessary to obtain deeper insights. We plan to carry out such a study in our future work. 

Although the use of the same best-performing prompt to generate explanations for examples of different difficulties was an important design decision to explore the generalizability of the approach, it might be possible that different prompts will perform best for examples of different difficulties. We will explore this opportunity in the next round of our work.

We also observed that for some lines of code in our dataset, experts, ChatGPT, or both choose to provide no explanations. In the current study, these lines were excluded from the evaluation since a meaningful comparison was not possible. However, choosing whether to explain a specific line or not is an important decision and the current study did not assess who is making better decisions about skipping lines, ChatGPT or experts. This aspect requires further investigation. In our next study, we plan to ask participant evaluators to specify whether each line of code needs an explanation or not. 

Another potential limitation of the study was the lack of a formal definition of what a ``complete'' explanation means during external evaluation. We let the participants decide how to rate completeness since it is a personal decision, which editors should make when deciding whether to update generated explanation or not. While it was a natural thing to do, it might have decreased the agreement between evaluators.  In our future work, we will see whether the agreement could be increased by defending correctness and completeness ratings more formally.

Finally, one aspect of human-AI collaboration not explored in this study is the value of keeping our engineered prompt open to the authors to change. Existing research reviewed above demonstrates that users unfamiliar with LLM are not able to produce well-performing prompts~\cite{10.1145/3544548.3581388}. However, most instructors and TAs in programming courses are computer scientists with graduate-level training. We expected that some fraction of these users could benefit from the ability to change the prompt and leave this option open. This assumption, however, has to be explored. We hope that a study that engages real instructors or TAs in producing worked examples for their course might provide interesting data on end-user work with a prompt. 
The ultimate way to address these limitations and collect valuable information is to run a multi-semester-long study engaging instructors to use the tool to produce explanations. Such a study will also enable us to assess the quality of explanations produced through human-AI collaboration and their value for students in introductory programming classes.

\bibliographystyle{ACM-Reference-Format}
\bibliography{references}

\end{document}